\documentclass[11pt,fleqn]{article}
\usepackage{graphicx,multicol}
\begin{document}
\centerline{\LARGE\sf Neutrinos from galactic sources of cosmic rays} 
\vskip1mm
\centerline{\LARGE\sf with known $\gamma$-ray spectra}
\vskip3mm
\centerline{Francesco Vissani}
\centerline{\em INFN, Laboratori Nazionali del Gran Sasso, Assergi (AQ)}
\vskip4mm
\noindent{\footnotesize We describe a simple procedure to estimate 
the high-energy neutrino flux from the observed $\gamma$-ray spectra 
of galactic cosmic ray sources that are transparent to their
gamma radiation. 
We evaluate in this way the neutrino flux from the supernova remnant 
RX J1713.7-3946, whose very high-energy 
$\gamma$-ray spectrum (assumed to be of hadronic origin) 
is not a power law distribution according to H.E.S.S.\ observations.
The corresponding muon signal in neutrino telescopes is found to be about 5 
events per km$^2$ per year in an ideal detector.}
\vskip1mm
\noindent{\footnotesize\em PACS: Neutrinos from CR 98.70.Sa;
Observations of $\gamma$-rays 95.85.Pw; 
SNR in Milky Way 98.38.Mz.}

\vskip2mm
\section{\sf Context and motivations}
The recent observations 
of $\gamma$-rays above 
TeV by H.E.S.S.\ are 
of great interest \cite{hessSelected,hess}. 
They will certainly help in answering 
the old question of the origin of the cosmic rays till the knee
\cite{berez,gaisser,stanev,aharonian,pt} and  
at the same time they  could provide 
us a reliable guidance for what we should expect in 
neutrino telescopes, at least for certain sources.

This is evident for the main candidate  
sources of galactic cosmic rays, supernova remnants (SNR) \cite{baade,ginz}.
The huge kinetic energy of the gas of the SNR could be effectively converted
into cosmic rays by diffusive shock acceleration \cite{dsa}, producing 
enough cosmic rays to compensate the losses from the Milky Way. 
When the SNR is surrounded by matter that can act 
as a target for cosmic rays, we would have a point source of 
very high-energy (VHE)
gamma radiation, which seems in agreement with certain
observations. Since we expect that the matter around SNR is 
not too dense anyway, the $\gamma$-rays are not significantly
absorbed, and there is a rather direct relation 
between VHE $\gamma$-rays and neutrinos. 

More in general, we think that it is important 
to take advantage of the detailed observations of $\gamma$-rays
whenever they exist in order to formulate definite 
expectations for neutrino telescopes.
This is certainly true after the most recent H.E.S.S.\ observations,
that are beginning to find VHE $\gamma$-ray spectra that deviate
from power law distributions above 10 TeV or so \cite{hess}.

Our recipe to calculate the neutrino fluxes  is described
in Sect.\ \ref{2} and the application to the SNR 
RX J1713.7-3946 is in Sect.\ \ref{3}. 
In essence, these results are a straightforward application of standard 
techniques \cite{lip} (and we follow as much as possible the
conventions of \cite{gaisser} to emphasise this fact) 
but we hope that they are useful in the present moment, when
the high-energy gamma astronomy is flourishing and 
the neutrino telescopes are finally becoming a reality.

\section{\sf Deriving the neutrino flux from the $\gamma$-ray flux\label{2}}
Let us assume that the VHE $\gamma$-ray flux 
$F_\gamma$ observed from a certain source is of hadronic origin 
and  that it is not significantly absorbed--the source is 
$\gamma$-transparent.\footnote{Therefore, this procedure is not of 
direct application for a number of possible galactic sources 
of neutrinos such as micro-quasars \cite{dist}
that are intrinsically non-transparent
or even for extragalactic sources since  
the IR photons background absorbs the VHE gammas above 
$\sim 10$ TeV; see also \cite{aaa}.} From 
the well-known 
relation $F_\gamma(E)=\int_E^\infty dE'\ 2\ F_{\pi^0}(E')/E'$
valid for high-energy $\gamma$-rays we find:
\begin{equation}
F_{\pi^0}(E)=-\frac{E}{2}\ \frac{dF_\gamma}{dE}
\end{equation}  
that implies that the $\gamma$-ray flux has to be strictly 
decreasing.
This equation, together with the approximate isospin-invariant distribution
of pions: 
\begin{equation}
F_\pi \equiv F_{\pi^0}
\approx F_{\pi^+}\approx F_{\pi^-}
\end{equation}
permits us to predict the flux of neutrinos 
using the observed $\gamma$-ray flux. 
It is important to note that the 
charge asymmetry has a small or negligible impact on the observable 
$\nu_\mu$ flux, compare \cite{cost,asimm}. 
The $\nu_\mu$ flux from the decay $\pi^+\to \mu^+ \nu_\mu$ is:
\begin{equation}
F_{\nu_\mu}(E) 
=\frac{F_\gamma(E/(1-r))}{2(1-r)} 
\label{nu1}
\end{equation}
where $r=(m_\mu/m_\pi)^2$.
The neutrinos from muon decay 
$\mu^+\to \bar{\nu}_\mu \nu_e  e^+$ have a more
implicit expression:
\begin{equation}
F_{\nu}(E_\nu)= 
\int^1_0 \frac{dy}{y}\ F_\mu(E_\mu)\ ( g_0(y)- \bar{P_\mu}(E_\mu)\  g_1(y) )
\label{nu2}
\end{equation}
where  $E_\mu = E_\nu/y$ and $g_i$ are known polynomials: 
$g_0=5/3-3 y^2+4/3 y^3$  
and $g_1=1/3-3 y^2+8/3 y^3$ when $\nu=\bar{\nu}_\mu$, while
$g_0=2-6 y^2+4 y^3$ and 
$g_1=-2+12 y -18 y^2+8 y^3$ when $\nu={\nu}_e$ \cite{lip}.
The muon flux (from $\pi^+$) that appears in previous formula is:
\begin{equation}
F_\mu(E) 
=\frac{F_\gamma(E)-F_\gamma(E/r) }{2(1-r)} 
\end{equation}
while the muon polarisation averaged over the pion distribution is
given by:
\begin{equation}
\bar{P_\mu}(E)\times F_\mu(E)  = -\frac{F_\gamma(E)+F_\gamma(E/r)}{2(1-r)}+
\frac{r}{(1-r)^2} \int_E^{E/r}\!\! F_\gamma(E')\ \frac{dE'}{E}
\end{equation}
It is easy to check that in the special case of power law distributions
these equations reproduce the results of Sect.~7.1 of \cite{gaisser}
(e.g., eq.~7.14 there).

We include the contribution to $\gamma$ flux from $\eta\to \gamma \gamma$ 
(resp., the contribution to $\nu$ flux from the leptonic $K^\pm$ decay)
in the simplest conceivable approximation: 
namely, we declare that the relevant flux of eta mesons (resp., the one of  
charged kaons) is proportional to the one of the neutral pions 
(resp., of the charged pions) with a fixed coefficient 
$f_{\eta}= 10$~\%  (resp.\ $f_K=25$ \% $\times 0.635$).
Thus: 
(1)~all formulae above should be multiplied by $1/(1+f_\eta)$, 
and then (2)~we  add 
a neutrino contribution that has the 
same form as the one from charged pions, 
but replacing $r = (m_\mu/m_K)^2$ and including 
the multiplicative factor $f_K$.

Finally, we incorporate 3 neutrino oscillations replacing:
\begin{equation}
F_{\nu_\mu} \to F_{\nu_\mu}\ P_{\mu\mu} +F_{\nu_e}\ P_{e\mu}
\label{oosc}
\end{equation}
and the same for antineutrinos. 
The numerical values of the oscillation probabilities are 
$P_{\mu\mu}=0.39\pm 0.05$ and $P_{e\mu}=0.22\mp 0.05$ where 
the quoted errors, approximately equal and opposite, 
are mostly (0.04) due to the spread of $\theta_{23}$ 
around maximal mixing 
and partly (0.02) to the spread
of $\theta_{13}$ around zero; 
the effect of the 
uncertainty in $\theta_{12}$ is smaller. 
See \cite{cost} for more discussion,
\cite{nudata} for a resum\'e{} of neutrino 
data and analysis, and \cite{rh} for further references.

We note in passing 
a stricter condition on the behaviour of 
the flux of VHE secondaries with the energy. Consider the
connection with the primary cosmic rays 
$F_\pi(E)\propto \int_E^\infty dE' F_p(E') k(E/E')/E $,
that we assume for simplicity to be protons.
When we go from $E=E_1$ to $E=E_2$ with 
$E_2>E_1$, the integral decreases because 1)~the lower limit increases;
2)~the scaling distribution $k$ in the integrand is a decreasing
function; 3)~there is an explicit factor $1/E$. 
Thus, also $f_\pi(E)\equiv E F_\pi(E)$ decreases. 
The same can be said of
the function $f_\gamma(E)=E F_\gamma(E)$, since 
$f_\gamma(E)=2\int_0^1 f_\pi(E/z) dz$; in other words the flux of 
hadronic $\gamma$-rays decreases {\em at least} as $1/E$ 
at high energies.\footnote{Such a 
very hard spectrum would follow from a hypothetical population
of very energetic primaries. In fact, consider $F_p(E')=\delta(E'-E_0)$:
when $E\ll E_0$ we find that the pions have $F_\pi(E)\propto 1/E$ 
since $k(0)\neq 0$; thus, the $\gamma-$rays would obey 
the $1/E$ distribution.}

\section{\sf Application: neutrinos 
from RX J1713.7-3946\label{3}}
\begin{figure}[t]
\caption{\em Expected $\nu_\mu$ fluxes corresponding to the two 
$\gamma$-ray spectra of eq.~\ref{2f}.\label{figfig}}
\centerline{\includegraphics[width=.52\textwidth,angle=270]{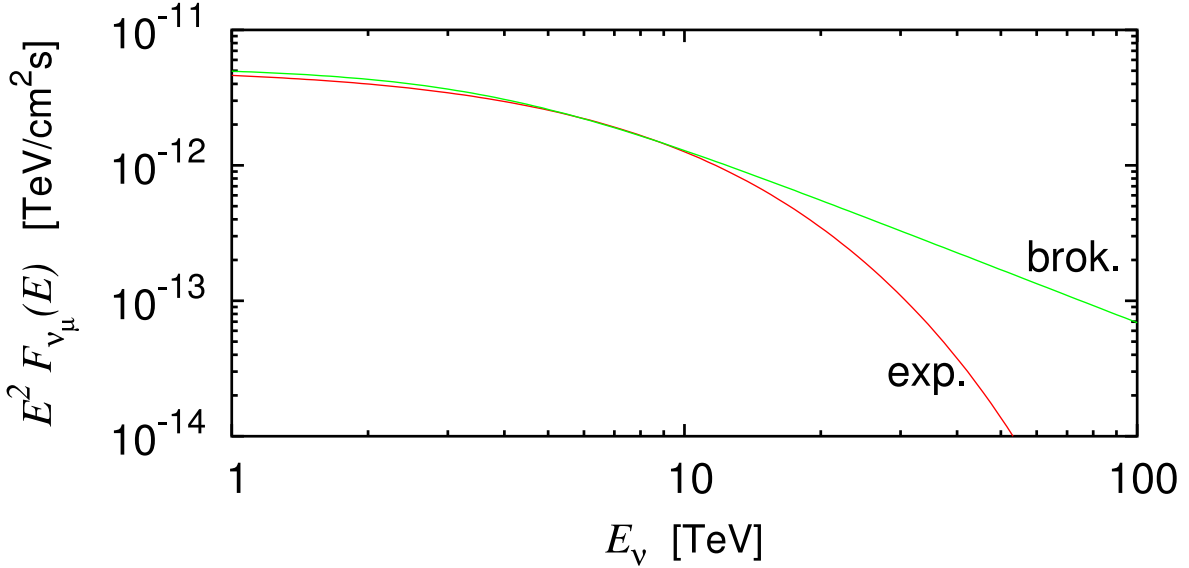}}
\end{figure}

We apply the formalism of the previous section to obtain the expected 
neutrino flux from RX J1713.7-3946 on the basis of H.E.S.S.\ observations
\cite{hess}. We use 2 parameterisations of the $\gamma$-ray flux
that describe well the observations \cite{hess}:
\begin{equation}
F_\gamma(E) = \left\{
\begin{array}{ll}
20.4\ E^{-1.98}\ \exp(-E/12) 
 & [\mbox{exponential cutoff}]\\[1ex]
20.1\ E^{-2.06}\
[1+ (E/6.7)^{2.5}]^{-0.496} &
[\mbox{broken power law}]
\end{array}
\right.
\label{2f}
\end{equation} 
where the units are TeV
for the energy $E$ and $10^{-12}/(\mbox{TeV cm$^2$ s})$ for the 
flux $F_\gamma(E)$. 
We do not use the third parametrization proposed 
in \cite{hess}, $F_\gamma\propto E^{-2.08- 0.3 \log E}$,
namely the distribution with energy dependent photon index:
in fact, this cannot result from $\pi^0$s,
since this is just a Gaussian in the 
logarithmic variable $\log E$, that increases 
rather than decreasing before 40 GeV. 
Note that a relatively low cutoff 
implied by H.E.S.S.\ observations is consistent 
with the present theoretical expectations \cite{bereze}
(however, rather different models of the same object 
are also discussed \cite{malk}).
The result for muon neutrinos
according to formulae \ref{nu1}, \ref{nu2} and \ref{oosc} 
is presented in figure \ref{figfig} and table \ref{tabtab};
in our approximation, the flux of antineutrinos is the same. 

We can estimate the number of through-going events 
in a neutrino telescope with $E_{th}=50$ GeV, $\phi=42^\circ 50'$ (ANTARES 
location) following  \cite{cost}. 
Considering neutrinos 
with energies below $E_{\nu,\rm max}=300$~TeV 
(that is not a significant limitation), 
we find:\footnote{The same numbers are obtained using
$F_{\nu_\mu}+F_{\bar{\nu}_\mu}=0.37 F_\gamma$.} 
\begin{equation}
N_{\mu+\bar{\mu}}=\left\{
\begin{array}{ll}
\mbox{4.8 per km$^2$ per year} &\mbox{ [exponential cutoff] }\\
\mbox{5.4 per km$^2$ per year}&\mbox{ [broken power law]}
\end{array}
\right.
\label{rrr}
\end{equation}
that can be compared with $N_{\mu+\bar{\mu}}= 8.8$ 
of \cite{cost}, obtained assuming a power law distribution. 
Thus, the new H.E.S.S.\ data suggest a signal about 
8 times weaker than given in~\cite{halzen} 
(resp., 2 times weaker than in \cite{cost}) 
that adopted a power law distribution 
extrapolated from the first observations of  CANGAROO
(resp., of H.E.S.S.).

One can gain something if some 
events above the horizon are accepted; e.g., with $5^\circ$  more,
one can go from a fraction of time useful for observation of 
78 \% (used for the numbers quoted in eq.~\ref{rrr}) 
to 88~\%{}. This is similar
to effects here neglected, e.g., other contributions of 
$\eta$ and $K$ meson decays or the deviations 
of $\theta_{23}$ from maximal mixing,
and should be comparable with the error of the method of calculation
we proposed. The effect of finite detection efficiency for 
realistic detector configurations instead should be more important 
(comparable with the effect of the deviation 
from a power law distribution discussed
here) for the events are not expected to be particularly energetic:
the distribution of parent neutrino energies 
has a median of 3 TeV for both distributions of eq.~\ref{2f}.

\begin{table}[t]
\begin{caption} {\em $1^{st}$ line: selected values of neutrino energy. 
$2^{st}$ line: sum of the yields of muons and antimuons (including 
Earth absorption), times the reference  
area $A=1$~km$^2$ and observation time $T=1$ year. 
$3^{nd}$ and $4^{rd}$ line: the $F_{\nu_\mu}/F_\gamma$ 
ratio for the $\gamma$-ray fluxes of eq.~\ref{2f}, 
which varies significantly with the energy.
\label{tabtab}}
\end{caption}
\begin{center}
\begin{tabular}{c|cccccccc}
$E$ [TeV]              & .1 & .3 & 1 & 3 & 10 & 30 & 100 & 300 \\  
\hline \\[-2ex]
$A T (Y_\mu \! + \! Y_{\bar{\mu}}  ) $ [cm$^2$ s]  
& {\small $ 1.0\mbox{\tt\scriptsize E}{9} $} 
& {\small $ 2.3\mbox{\tt\scriptsize E}{10} $} 
& {\small $ 2.8\mbox{\tt\scriptsize E}{11} $} 
& {\small $ 1.9\mbox{\tt\scriptsize E}{12} $} 
& {\small $ 1.1\mbox{\tt\scriptsize E}{13} $} 
& {\small $ 3.5\mbox{\tt\scriptsize E}{13} $} 
& {\small $ 9.2\mbox{\tt\scriptsize E}{12} $} 
& {\small $ 1.6\mbox{\tt\scriptsize E}{14} $} \\
$F_{\nu_\mu}/F_\gamma$, exp.~cutoff& .26 & .26 & .25 & .21 & .14 & .06 & .02 & .01 \\  
$F_{\nu_\mu}/F_\gamma$, brok.~power& .25 & .25 & .25 & .21 & .14 & .13 & .13 & .13 \\
\end{tabular}
\end{center}
\end{table}

\section{\sf Summary and discussion}
In summary, we presented a simple procedure to convert the observations
of high-energy $\gamma$-rays into expectations for high-energy neutrinos,
assuming that the source is gamma-transparent and that the flux
of VHE
$\gamma$-ray is due to cosmic ray interactions (=it is of hadronic 
origin). The latter hypothesis shows that our flux should be thought 
as an upper bound for gamma-transparent sources. 
As an application, we calculated the neutrino flux from RX J1713.7-3946 
expected on the basis of new H.E.S.S.\ results and found that 
the expected number of events decreases by $40-50$~\% 
and that the signal consists of relatively low energy
events.

In the future, it will be interesting to 
repeat the same steps for other intense sources of VHE $\gamma$-rays, 
e.g., Vela Jr (RX J0852.0-4652), that is almost 
continuously visible from ANTARES (95~\% of time).
In the region $E\le 10$~TeV \cite{hessSelected}
the spectrum is described by $F_\gamma=21\ E^{-2.1}$ 
(same units as in eq.~\ref{2f}).
Suppose that the future observations will demonstrate 
a milder exponential cutoff, described by a multiplicative 
factor $\exp(-E/E_{\rm cut})$ with $E_{\rm cut}=50\ (150)$ TeV. 
In this case we 
would find $N_{\mu+\bar{\mu}}=10$ (14) events per km$^2$ per year 
with a median neutrino energy of 5.5 (8.5) TeV (if, again, 
we assume that all $\gamma$-rays are hadronic). 
If instead RX J1713.7-3946 should turn out to 
represent a typical SNR in a typical stage, it will be important   
to understand the cosmic rays from a few hundred TeV till the knee, 
e.g., considering other galactic point sources of cosmic rays 
and/or further phases of cosmic ray acceleration.

These results emphasise even further the importance 
to obtain $\gamma$-ray 
observations in the region from 10 to 100 TeV
and to understand well the experimental background coming from atmospheric 
neutrinos.

\vskip4mm
\noindent We gratefully thank 
F.~Aharonian,
V.~Berezinsky,
P.L.~Ghia, 
D.~Grasso and especially 
P.~Lipari  
for useful discussions and 
M.L.~Costantini for help.

\section*{Note added} After this work  
was submitted for publication and when it was presented 
at Vulcano 2006 conference (May 2006), 
a number of interesting new works appeared: \cite{lipari} 
where the background and possible strategies for neutrino search
are quantitatively discussed; \cite{aro}, where a 
detailed parameterizations of neutrino and gamma yields is offered;
\cite{uli}, where neutrinos events from Vela Jr are 
estimated (though, without describing the details of the 
calculation) using a fixed $\nu/\gamma$ 
conversion coefficient $=1/2$.

\begin{footnotesize}
 \def\refname

\end{footnotesize}

\end{document}